\documentstyle[12pt]{report}
\begin{document}
\hoffset=.75in
\voffset=.25in
\begin {titlepage}
\begin{flushright}
SU-4228-498

TPI-MINN-91/39-T

January, 1992
\end{flushright}
\vspace{2 in}
\centerline {\bf {\LARGE Topological Spin-Statistics Theorems for Strings}}

\begin{center}
A.P.Balachandran*,  W.D.McGlinn $^{\dagger}$,  L.O'Raifeartaigh
$^{\ddagger}$,

 S.Sen $^{\sharp}$,  R.D.Sorkin*,  A.M.Srivastava**

 {\sl *Physics Department, Syracuse University, Syracuse,NY 13244-1130,USA

 $^{\dagger}$ Physics Department, University of Notre Dame,Notre Dame,IN
46556,USA

 $^{\ddagger}$ Dublin Institute for Advanced Studies,10 Burlington
Road,Dublin 4,Ireland

 $^{\sharp}$ School of Mathematics, Trinity College, Dublin 2, Ireland

 **Theoretical Physics Institute, University of Minnesota,Minneapolis,

 MN 55455,USA }

 \end{center}

 \noindent{\bf Abstract:} Recently, a topological proof of the
spin-statistics theorem
 has been proposed for a system of point particles. It does not require
relativity
 or field theory, but assumes the existence of antiparticles. We extend
this proof
to a system of string loops in three space dimensions and show that by
assuming the
existence of antistring loops, one can prove a spin-statistics theorem for
these string
 loops.  According to this theorem, all unparametrized strings (such as
flux tubes
 in superconductors and cosmic strings)\ should be quantized as bosons.
Also, as in the
 point particle case, we find that the theorem excludes nonabelian
statistics.

 \end{titlepage}

It is a general expectation that point particles as well as solitonic
excitations should obey the standard spin-statistics connection. In three
or more
spatial dimensions, this amounts
to saying that objects that are characterized by tensorial states
(i.e. states which are invariant under 2$\pi$ rotations) should obey Bose
statistics and objects characterized by spinorial states (i.e. those
which change sign under 2$\pi$ rotations) should obey Fermi statistics.
Statistics being related to an exchange of two identical objects,
the spin-statistics connection thus asserts that the change of phase of a
state
under an exchange of two identical objects of spin S is
exp[i2$\pi$S]. This correlation is also expected to be true for
objects in two space dimensions where S can be any real number.

  No experimental evidence exists which indicates a violation of this
   spin-statistics connection. From
the theoretical side, there are two, essentially different, classes
of proofs that have been put forward. One class of proofs occurs in
the context of relativistic quantum field theory where it is
shown that tensorial fields commute and spinorial ones anticommute
for space like separations$^{1}$. The second approach is topological
in nature and originates from the work of Finkelstein and
Rubinstein$^{2}$ on certain field theory solitons. [Extension of this work
to cover
a more general class of field theory solitons was carried out by Friedman
and Sorkin $^{3}$ and by Sorkin$^{4}$.]
This approach does not require assumptions
like locality, Poincare invariance etc. which were needed in the
former class of proofs. It examines the fundamental group $\pi_1$(Q)
of the configuration space Q appropriate for solitons and shows that the
2$\pi$ rotation of a soliton and the exchange of two identical solitons
are the same elements of $\pi_1$(Q) [or as we shall sometimes say,
"equivalent"] and hence must have the same
action on the states.

 It has been emphasized  in refs.4 and 5 that the existence of antisolitons
with which solitons can be annihilated and be pair produced is
of crucial importance in these topological approaches.
Following this line of thought, the
topological approach was extended in ref.6 to the case of point particles
by examining the configuration space of a system of identical
particles and their antiparticles. Such a proof for the spin-statistics
connection is of great relevance especially for certain condensed
matter systems where particle-hole pairs fit the rather  general
definition of the particle-antiparticle pairs used in ref.6. It is
important to note that none of the earlier proofs apply to these
cases of particle-hole excitations.

  In this paper, we further this line of work by constructing a
topological proof for the spin-statistics connection for a system of
loops of strings. The strings we are interested in are not the
fundamental strings. We do not require the invariance of
the action under reparameterisations of the world sheet
coordinates or under the  Poincar\'{e} group. Hence there is no compelling
theoretical reason for us to insist that the space-time
dimension differs from four.
There are numerous  physical systems
where such strings occur, examples being vortices in superfluid
helium, flux tubes in type II superconductors and strings formed in
the early universe due to a Grand Unified Theory phase transition.
Our approach is based on methods used in ref.6
for the case of point particles. We find that essentially the same
set of ingredients as those required for the point particle
case work for the string case as well. This shows to us
that these proofs of the spin-statistics connection
(those constructed here and in ref.6) are of a very general nature and
should apply
to objects in a vast class of models.

  The configuration space of a system of loops of strings
has been previously investigated in ref.7 where many unusual
possibilities for statistics were found. For example, it
was shown in ref.7 that one could quantize a system of string loops
such that the strings do not have any definite statistics in the sense
that their states are not eigenstates of
the exchange operator. It was also possible to quantize
tensorial strings (for which 2$\pi$ rotation is trivial) as
fermions. We will show in this paper that once antistrings
are introduced and possibilities of string-antistring pair
production and pair annihilation are allowed, a
spin-statistics connection can be established for strings
and all of the above mentioned exotic possibilities can be eliminated.
We should point out that, as in ref.7, we do not allow the string loops
to shrink down to or emerge from the vacuum.  This can be achieved if
strings carry some conserved charge.  The antistrings will then carry the
negative charge.

 Let us start by  summarizing the essential ingredients in the
construction of the topological proof of the spin-statistics
connection for point particles $^{6}$. As we indicated
earlier, the proof consists in establishing a homotopy between the loop
corresponding
to the 2$\pi$
rotation loop of a particle and the loop corresponding to the exchange
of two identical particles. The full configuration space is constructed
from the configuration spaces of identical particles and the configuration
spaces of identical antiparticles and by prescribing rules for the
annihilation and  production of particle-antiparticle pairs. Here the
configuration space of (say) particles (with no spin) consists in
specifying
the locations of particles and by identifying two configurations which
differ by a permutation of particle locations. No two particles are allowed
to coincide in space. The configuration space with inclusion of spin (for
three
 space dimensions) is constructed by attaching orthonormal, right handed,
3-frames
to particles and orthonormal, left handed, 3-frames to antiparticles in the
 preceding space. Specific handedness of the frames is a matter of
convention, but the relative
handedness of particle and antiparticle frames is dictated by the
fact that CP or CPT transforms a left handed particle into a
right handed antiparticle. It also follows from the examination of
soliton and antisoliton configurations. [See ref.6 for more
details.]

  It was found necessary in ref.6 to impose two additional conditions
concerning
 the processes of pair creation and pair annihilation. The first
one, termed the ``mirror condition'', requires that a
particle-antiparticle pair annihilates to the vacuum (or be pair produced
from the vacuum) only if their frames are mirror reflections of
each other in a plane which is perpendicular to the direction
of approach (or the direction of separation). A justification of this
mirror
condition can be obtained by analysing soliton-
antisoliton annihilation (or creation) processes $^{6}$. The second
condition (``syzygy condition'') requires that two
particles and one antiparticle (or two antiparticles and a particle)
 can simultaneously overlap only if
they become colinear as they approach the point of overlap.
The justification of this condition follows from the requirement
that the topology of the configuration space be Hausdorff. It follows
as well from soliton physics$^{6}$.

  We now proceed with the construction of the proof for the case
of strings. Both conditions mentioned above, namely,the
mirror and syzygy conditions, will be necessary
for  the proof.  They can be justified
 along  lines similar to those in ref.6. We may mention
here that for the point particle case$^{6}$, it was further shown that
the configuration space admits a topology which leads to the mirror and
syzygy
conditions. Such a verification is necessary since these
configuration spaces are very complicated and their continuity
properties are far from  evident. For the string case
however, we will not attempt to formally define a topology here, leaving it
for
a future work.

  We shall consider strings which are closed non-intersecting loops
classified in the
following categories.

 (a) Parametrized strings without frames.

 (b) Unparametrized strings with or without the specification of
an orthonormal transverse 2-frame at each point of the strings. We will
also assume
that the frames have a net zero `winding ' number around the string.

 (c) Parametrized strings with frames of the above sort.

 The above classification of strings is dictated by their
properties under 2$\pi$ rotation and under exchange. We will
show that unparametrized strings, with or without frames, (class (b))  are
tensorial
in the sense that 2$\pi$ rotation is (homotopically) trivial for them. This
is actually
 rather clear
since one can use as a base point for  homotopy a planar circular loop
with, for the
framed case, the 2-frame at each point having one unit vector
lying in the plane of the loop. It is then clear that a rotation of the
string in the
 plane of the loop is trivial.  We will also prove that the exchange loop
is equivalent
 to this rotation and is thus trivial as well.  Hence
 these strings are bosons in accordance with
the spin-statistics connection. We may mention
here that most commonly discussed strings such as vortices in
superfluid helium, flux tubes in type II superconductors and cosmic
strings belong to class (b) and therefore should be bosons.
They must of course carry some conserved charge  in
order to prevent their collapse.

 We will show  that  strings in class (a)  should
be considered as spinless strings since the exchange loop
of two such identical strings, which is homotopic to the 2$\pi$ rotation of
 one string, is homotopically trivial. [This is so even though {\em in the
 absence of creation-annihilation processes}, the 2$\pi$ rotation of a
single string
 is homotopically nontrivial and hence admits quantizations
  which violate the spin-statistics theorem. ]
 These strings  should not thus
be considered as proper analogues of particles with spin. This is to be
expected
from the point particle case$^{6}$ since particles with spin
are described by attaching orthonormal 3-frames to the
particle locations whereas class(a) strings possess no such frames.
 Since  strings are not rigid however, one can
not attach a fixed 3-frame to a string. The appropriate
analogue of particles with 3-frames is realized by strings in
class (c) where the 2-frames along with the parametrization
provide a proper generalization of 3-frames of particle mechanics.

  Let us first consider the configuration space $Q_M$(Par) of
M identical parametrized loops. The formal
definition of a single  parametrized string loop is as follows.
Let S$^1$ be a circle parametrized by the variable
$\sigma$ with the identification of $\sigma$ and
$\sigma + 2\pi$. A parametrized string in R$^3$ is then a
one-to-one smooth map  $x = (x^1, x^2, x^3)$ of S$^1$ into R$^3$:

 $$ x : S^1 \rightarrow R^3,$$
 $$ \sigma \rightarrow x(\sigma) = (x^1(\sigma),
     x^2(\sigma), x^3(\sigma)).\eqno(1)$$

 For $M > 0$, $Q_M(Par)$ is then defined as

 $$ Q_M(Par) = \{[x^{(1)},x^{(2)},..,x^{(M)}];$$
 $$  [x^{(1)},x^{(2)},..,x^{(i)},..,x^{(j)},..,x^{(M)}] =
[x^{(1)},x^{(2)},..,x^{(j)},..,x^{(i)},..,x^{(M)}]\}.\eqno(2)$$

\noindent where $x^{(i)}(\sigma)\neq x^{(j)}(\sigma^{\prime})$
 for $i \neq j$ or $\sigma\neq\sigma^{\prime}$  .
 This assumption that  strings do  not intersect each other or
 self-intersect is similar to the assumption that no two identical
particles occupy the same position in the identical particle configuration
space.

 The configuration space of N identical, parametrized, antistrings
is constructed in the same way. For notational convenience, we will denote
individual antistrings by barring $x$'s and their configuration space by
$\bar Q_N$(Par).

 The analogue of the mirror condition for parametrized strings (without
2-frames at each point) is
the condition that a string and antistring can touch at a point
only if their parametrizations agree at that point. Thus
$x^{(i)}(\sigma) = \bar x^{(j)}(\sigma^\prime)$ only if
$\sigma = \sigma^\prime$.
For a justification of this condition, we can appeal to known
physical examples of parametrized strings. For example, baryons in
the string picture of QCD can be considered as a triangle of strings
with quarks of three different colors at the vertices of this triangle.
These quarks can then be thought of as providing marks on
this baryonic string loop. If  one now considers an antibaryonic
loop (with antiquarks at the vertices), we see that annihilation is
possible only if a given quark is matched with the antiquark of
the same color, that is only if the  marks are of the same color.  The
similarity of this
rule of annihilation and the one we impose should be evident.
 We should emphasize here that this discussion (as well
as the justifications provided for mirror condition etc.) are only
meant to provide physical examples where our choices are realized.
We take these conditions as assumptions which are necessary to establish
a spin-statistics connection.

   We can now define the full configuration space $C_K$(Par)
appropriate for parametrized strings and antistrings with net string number
equal to  K when creation and annihilation processes are permitted. It is

  $$C_K(Par) =\begin{array}[t]{c}\bigcup \\[-.10in]{\scriptstyle{N+K \ge 0,
N \ge 0}}\end{array}
     Q_{N+K}(Par)\times \bar Q_N(Par) / \sim\eqno(3)$$

\noindent where $Q_0(Par) \times \bar Q_0(Par)$ is defined as the vacuum.
$\sim$ is the equivalence relation which accounts for pair creation
and pair annihilation and is defined by the following identifications:

 $$([x^{(1)},x^{(2)},..,x^{(i)},..,x^{(M)}] ;
[\bar x^{(1)},\bar x^{(2)},..,\bar x^{(j)},...,\bar x^{(M)}])$$
 $$\sim ([x^{(1)},x^{(2)},..,\underline{x}^{(i)},..,x^{(M)}] ;
[\bar x^{(1)},\bar x^{(2)},..,\underline{\bar x}^{(j)}},...,\bar
x^{(M)}])$$
 $$ if \ x^{(i)}(\sigma) = \bar x^{(j)}(\sigma) \ for \ all \ \sigma
.\eqno(4)$$

\noindent Here the underlined entries are to be deleted. The condition
$ x^{(i)}(\sigma) = \bar x^{(j)}(\sigma)$ for all
$\sigma$ refers to the situation
shown in Fig. 1A where the string and the antistring have completely
overlapped. This is the situation of complete annihilation of the string-
antistring pair and will be thought of as a vacuum configuration. It is
also possible that the string and antistring overlap in a small
segment which is annihilated as shown in Fig. 1B. This configuration is
not yet a vacuum configuration. It is identified with  any
 configuration of a string-antistring pair obtained by adjoining the
`missing'
  string - antistring sections to Fig.1B so that they overlap.
  (This is allowed when the $\sigma$'s appropriately match). Thus  Fig.1C,
for example, is identified with
  Fig.1B. A sequence of such configurations with more and more overlap
leads to complete annihilation and hence to the vacuum.

  We are now set to prove that the exchange loop for these strings is
trivial.
  The exchange we consider will be such that the
  string and antistring lie in the same plane (i.e. the x-y plane)
  in the initial configuration and throughout the exchange path.
The exchange path is depicted in Fig. 2 .  The homotopy parameter $\tau$
evolves
  vertically up and the $\sigma$ = 0 `world lines'  are indicated by the
thick lines.
The path in configuration space of Fig.  2  can be smoothly deformed
 into Fig.  3    in which string creation and annihilation  occur.
[Following Feynman, we
 adopt the convention that a string travelling backwards in `time' $\tau$
is an
antistring travelling forward in $\tau$.]
  The string- antistring configuration  for $\tau$  slices corresponding
  to  A,B,C,...,F are drawn in Figs.4     A, B, C, ....F.  In these
figures, antistrings
   are drawn with thick lines and the positions of
    $\sigma$ = 0  are marked with short lines.  The middle configuration in
Fig. 4 B is
    part string and part antistring. Note that the $\sigma$ range of
   the string part agrees with the $\sigma$ range of the antistring part.
    Fig. 3 can be continuously deformed into Fig. 5. The
    intermediate configuration at $\tau$ parameter
 D is the same as in Fig. 4 D.  Fig. 5 can be continuously deformed into
Fig. 6.

We will first show  that the path of the left string can be deformed
continuously to
 a 2$\pi$ rotation of the string, and then show that it can be deformed to
the trivial path,
 thus showing that both the exchange  and  2$\pi$ rotation paths of
 the string are trivial.   First consider configurations for  $\tau$ slices
corresponding
  to A,B,...,E which are drawn in Fig. 7.
We will argue that this path in configuration space
is homotopically equivalent to a 2$\pi$ rotation of the string about an
axis perpendicular
to its plane.  Note that in the sequence depicted,
 string 2$'$  must move off the x axis to allow the  string  1 to
annihilate
with antistring.  To bring the string - antistring creation and the
string-antistring
annihilation
together, so that there is no annihilation or creation process, one has to
deform the
path in
configuration space so that  string 2$'$ is created to the
right of the antistring.    Consider the string-antistring configuration
that emerges
 if the production angle $\theta$ is changed without changing the position
of the
  antistring.  This
  is depicted in Fig.8. Clearly, if one deforms the path so that the string
is emitted
   at an angle $\theta$, one must follow the emission with a - 2$\theta$
rotation about the z
   axis to
   bring the string back to its original angular position.  In particular,
if the
string is emitted at an angle of $\pi$, it must be followed by a 2$\pi$
rotation after
emission.
 Once the original path is deformed so that the emission angle is  $\pi$
and the emission is
 followed by a 2$\pi$ rotation, it can be further deformed so that  there
is no
string- antistring production and annihilation. The required
sequence of deformations, in which the creation process takes place
progressively later and the
annihilation process takes place progressively earlier in  $\tau$,
is indicated in Fig. 9.  Note that in this sequence, the final string
rotates by 2$\pi$.
Thus we have argued that the exchange path of Fig.2 is homotopically
equivalent to Fig.10
 in which  the left string undergoes a 2$\pi$ rotation about the z axis.

 We now argue that  the path of Fig. 6 can be deformed
into the trivial path. Suppose that immediately before the string-
antistring
configuration splits (see Fig. 7B), we distort
 the path so that the  part of the string which has yet to be created pops
off the `top'
(in the + z direction) of the part of the antistring
 which also has yet to be created  so that the  configuration
 depicted in Fig. 11 is obtained.
 (Fig. 11A depicts the configuration from above and Fig. 11B from the
side).  One
 then has to rotate the small part of the string about the y axis by an
angle
$\pm\pi$  after creation to restore the string to the circular state as
indicated in Fig. 11C.
If one deforms the path so that this occurs earlier in the creation
process,
one obtains  Fig. 12
.  Eventually the whole string 2$'$ is rotated by $\pm\pi$  after creation.
Since we want to eventually bring the creation and  the annihilation
processes together, we
 have to distort the path of  incoming string 1 so
  that it annihilates from the `bottom' (- z direction) of the
 antistring. Thus for the first small distortion, a small part of
 string 1 is rotated by $\pm\pi$    as depicted in  Fig 13A  and proceeds
to annihilate the antistring
 as depicted in Figs. 13B and 13C.

We can continue to distort the path until all of string 1 is rotated by
$\pm\pi$ about the
 y axis before being brought in to annihilate.  One now has a path in which
the string is
 rotated by $\pm\pi$ both before creation and after annihilation and string
2$'$ is created
above the antistring and string 1 is annihilated from below the antistring.
 The two
 processes  may now be merged so that no annihilation or creation takes
place, but rather
 a full 2$\pi$ rotation about the y axis or no rotation at all takes place.
 Thus we see
in this case the exchange and 2$\pi$ rotation are both
 homotopically equivalent to the trivial path.

Let us now discuss the case of unparametrized strings (class (b)). The
definition
of unparametrized strings follows from eqn.(1). The group D of
diffeomorphisms $\sigma \rightarrow \phi(\sigma)$ of S$^1$ acts on a map
$x$
according to the rule

 $$(\phi^*x)(\sigma) = x(\phi(\sigma)).\eqno(5)$$

 An unoriented, unparametrized string is the equivalence class $<x>$ of
all parametrized strings related to the parametrized string $x$ by the
action of
the full diffeomorphism group D. An oriented, unparametrized string is
the equivalence class $<x>$ of all parametrized strings related to the
parametrized string $x$ by the action of D$_0$ which is the identity
component of D. D$_0$ consists of all diffeos which do not reverse the
orientation of S$^1$.

Let us first prove that the 2$\pi$ rotation loop for unparametrized
(oriented or unoriented) strings about any axis is homotopically trivial
in the single string configuration space Q$_1$.
For this purpose, it is convenient to take the base point $<x_0>$
for $\pi_1(Q_1)$ to be a circular string in the x-y plane with origin as
center. Let the 3 $\times$ 3 orthogonal matrix R$_{\vec n}(\phi)$ denote
rotation by angle $\phi$ about axis $\vec n$. Then since R$_{\vec n}(0)$ =
R$_{\vec n}(2\pi)$ = 1,
the curve \{$<R_{\vec n}(\phi)x_0 >| 0 \le \phi \le 2\pi \}$ is a loop
for each $\vec n$. It describes 2$\pi$ rotation  of $<x_0>$ about $\vec n$.
Since $<R_{0,0,1}(\phi)x_0> = <x_0>$ for the above choice of $x_0$, the
variation of $\vec n$ from $\vec {n}_0$ to (0,0,1) provides a homotopy of
2$\pi$ rotation about $\vec {n}_0$
to a single point and establishes the result. [It is important to note here

that the above argument
can not be carried out for  parametrized strings and 2$\pi$ rotation
of parametrized strings in one string  configuration space is nontrivial.]

 Consider now the
exchange of two identical unparametrized strings (described by a figure
which looks
the same as the one shown in Fig. 2
for the parametrized case). We can follow all the steps shown from Fig. 2
up to
Fig. 10 as all these steps can be carried out for unparametrized strings
as well. With the base point  being two circular strings
 in the x-y plane, Fig. 10 is a trivial path.
Thus we see that  exchange is trivial for unparametrized strings
and that these strings are bosons in accordance with the spin-statistics
connection.

 Let us now discuss parametrized, framed strings (class (c)). For us, a
frame is
 an orthonormal 2 - frame attached to the string at $x(\sigma)$ and with axes
 normal to the tangent $x'(\sigma)$. The tangent vector at $\sigma$
 along with this 2-frame at
$\sigma$ thus provide an orthonormal 3-frame at each point of
the string. We can characterize this 2-frame by a unit vector
$\hat{n}(\sigma)$
at each point of the string orthogonal to the tangent vector at $\sigma$.
The configuration space $C_K^F$(Par) appropriate for  these
parametrized, framed strings and antistrings with net string number equal
to
K  is constructed from $C_K$(Par)
by replacing   $x^{(i)}(\sigma)$'s by
 $\{x^{(i)}(\sigma), \hat{n}(\sigma)\}$.  Further, we assume that our
configuration
 space is restricted to zero  winding number for $\hat{n}(\sigma)$, that is
that
 that $\hat{n}(\sigma)$ does not wind around the string with nonzero
winding number
 when $\sigma$ increases by 2$\pi$.
 As one can expect from our discussion of the point particle
case, the 2-frames for antistrings are then chosen so that the 3-
frame at each $\sigma$ of the antistring
has opposite orientation to the corresponding 3-frame
for the string. We emphasize here again that if the strings had rigid
shapes, prescribing a single 3-frame at the center of mass would have
been sufficient. However, physical strings can change their shapes and
it is therefore necessary to attach a 2-frame at each point of the string.

 Rules of annihilation (or production) of string-antistring
pairs now require  conditions for an appropriate matching of
$\hat{n}(\sigma)$ in addition to the condition that the parameter $\sigma$
should match at the point of contact (see eqn. (4)).
The analogue of the mirror condition for the point particle case dictates
here that as $x(\sigma)$ approaches $\bar x(\sigma)$ (as we mentioned,
$\sigma$'s must match at the point of contact), the unit vector of the
string and that of the antistring at point $\sigma$ must be mirror
reflections
of each other in the plane perpendicular to the direction along which
$x(\sigma)$ approaches $\bar x(\sigma)$.

 We will again show that the exchange of two framed strings is
 homotopic to the rotation of a single string.
 Similarly, we will see that since now there
are  unit vectors at each point of the strings, the  mirror
condition makes inapplicable those arguments  used to show
that the exchange loop is trivial for parametrized  strings.
We had already mentioned that the 2$\pi$ rotation loop is nontrivial
 for parametrized  strings. The same
is clearly true for parametrized framed strings as well. These strings thus
may represent a proper string
analogue of spinning particles. Of course the hole in the argument is that
there may
 be a deformation other than that used for the unframed strings which would
show
that the exchange is trivial for class (c) as well.  We will now show that
the exchange of
two such strings is homotopic to a 2$\pi$ rotation of a single string.

 Just as for unframed strings, we first consider the exchange of planar
 strings.  We assume also that all the unit vectors defining the frames lie
in the
  plane of the strings and point inward. Since we are restricting our
configuration
  space to consist only of strings with zero winding number, this is no
restriction.
   The initial configuration is depicted in Fig.14.
  When the string - antistring pair is created in a plane or annihilated
  in a plane,  the mirror rule allows us to assume  that the resulting
configurations are
   all planar,or more precisely that the  $\hat{n}(\sigma)$'s and
    $\hat{\bar{n}}(\sigma)$'s all lie in the plane of the string
    and antistring and point inward .  The
arguments for the case of production and annihilation in the plane,
illustrated by Figs. 2 to 10, go through unchanged and exchange is
equivalent
 to  a  2$\pi$ rotation of one of the strings as depicted in Fig.10. Note
that for framed
 unparameterized strings,  this is a trivial path so that the exchange path
is again trivial.

 However, the arguments change for non-planar production and annihilation
depicted
 in Figs. 11, 12, and 13.  Now, immediately before the string - antistring
configuration
  splits (Fig.7B), we cannot distort the path so that the  part of the
string which has yet to be created  pops off the top of that part of the
antistring
which has yet to be created.   This would lead to a configuration of string

2$'$ as depicted in Fig.15 which  is not allowed since  $\hat{n}(\sigma)$
is not smooth.
 Fig. 16   indicates how  one
can distort  the configuration of  string 2$'$ near the contact point with
the antistring
so that the uncreated part of the string can pop off
 the top of the uncreated part of the antistring.
The thick line represents part of the antistring and the light line
represents part of
string 2$'$.  After the rest of string 2$'$ is created, its configuration
is as depicted in Fig.17.
All of the  $\hat{n}(\sigma)$   are in the plane or above the plane. To
smoothly deform this configuration
to a planar loop with  $\hat{n}(\sigma)$   in the plane and pointing
inward, the small section of the string with outward directed
$\hat{n}(\sigma)$
    has to be rotated by an angle $\pi$ in the left handed sense around the
 y axis.
    If the section is rotated in the right handed sense around the y axis
    by an angle  $\pi$,  two twists of 2$\pi$
  of $\hat{n}(\sigma)$  would develop which would have to be unwound.
If one deforms the path so that this occurs earlier in the creation
process, one obtains a
configuration similar to the above, except that now the large portion  of
the string with the
 inward directed $\hat{n}(\sigma)$  has popped off the top of the
antistring and must be
 rotated by an angle $\pi$ in the left handed sense around the  y axis.
   Eventually the entire string 2$'$ must be rotated
by an angle $\pi$ in the left handed sense around the  y axis after
creation.
Since we want to eventually distort the path so as to bring the creation
process together
 with the annihilation process, we have to distort the path of the incoming

string so that it annihilates from the bottom of the antistring.  Thus a
small portion of
string 1 has to be rotated by an angle $\pi$ in the
left handed sense around the  y axis as shown in Fig. 13A, so that
 all \hat{n}($\sigma$)  are in the
plane or below  the plane.  We can continue to distort the path until all
 of string 1 is rotated  by an angle $\pi$ in the left handed sense around
the  y axis before
annihilation.  One now has a path in which  string 2$'$ is
rotated by + $\pi$ in the left handed sense around the y axis after
creation above the
 antistring and string 1 is rotated by + $\pi$ in the
 same sense before annihilation from below the antistring.  The
 two processes can now be merged so that no annihilaton or
creation takes place, but rather a full 2$\pi$ rotation about the y axis of
the string.
Furthermore we see that the sense of the rotations before and after
annihilation are fixed,
in contrast to what is possible for the unframed string, and thus the
exchange and 2$\pi$
rotations cannot be trivialised.

It is easy to see that our theorem excludes the
 possibility of nonabelian statistics for strings.
 Thus the exchange of any two strings in a system of several strings and
 antistrings is homotopic to a 2$\pi$
 rotation of one of the strings. An argument described in the first two
papers of ref.6
  then shows that exchanges, regarded as elements of the fundamental group,
commute,
  and nonabelian statistics is excluded.

 As  mentioned earlier, one needs to show that a suitable
topology can be defined on the configuration spaces appropriate
for various cases discussed here and that this topology implies
conditions such as the mirror and  syzygy conditions. We would like to show
this
in a future paper. With such a topology, one should then exhibit the
homotopy equivalence of the planar creation and the
 creation off the top of the antistring plus a $\pi$ rotation.
Also, without a careful definition
of the topology, and a more rigorous study of the spaces, one cannot
be sure that one is correctly deducing that the exchange is not trivial
even for framed strings.
After all, there may be a deformation of the exchange path to the trivial
path that we have overlooked.
 We conclude emphasizing again that one of the most important
 aspects of this topological proof of the spin-statistics theorem is the
fact that the set of ingredients  required here are almost identical
to those required in the widely different case of point particles
discussed in ref.6.

\noindent{\bf Acknowledgments}

The authors thank D. O'Connor for extensive discussions. The work of A.M.S.
was supported by the Theoretical Physics Institute at the University of
Minnesota and by the U.S, Department of Energy under contract number
DE-AC02-83ER40105.
The other authors were partially supported by National Science foundation
Grant
No.INT-8814944.  In addition, A.P.B. was partially supported by
the Department of Energy under contract No. DE-FG02-85ER40231 and R.D.S.
 by National Science Foundation under contract No. NSF PHY-9005790.
\newpage

\centerline{$\underline{\it REFERENCES}$}

\begin{enumerate}
 \item Cf. R.F. Streater and A.S. Wightman, {\sl P C T, Spin and
Statistics, and All That}  (W.A.
Benjamin,Inc., 1984); U.H.Niederer and L. O'Raifeartaigh, {\sl Fortschritte
der
 Physik} {\bf 22} (1974) 111 and 131, and references therein.
\item D.Finkelstein and J.Rubinstein, {\sl J. Math. Phys.} {\bf 9} (1968)
1762.
\item J.L.Friedman and R.D.Sorkin, {\sl Commun. Math. Phys.} {\bf 89}
(1983) 483 and 501.
\item  R.D.Sorkin, {\sl Commun. Math. Phys.} {\bf 115} (1988) 421.
 See also R.P. Feynman, {\sl The Reason for Antiparticles} in {\sl
Elementary Particles and the
 Laws of Physics} (Oxford University Press, 1987)
\item R.D. Tscheuschner, {\sl Int.J. Theor. Phys. } {\bf 28} (1989) 1269;
Hamburg preprint(1990).
\item A.P. Balachandran, A. Daughton, Z.-C. Gu, G.Marmo, R.D.Sorkin and
A.M.Srivastava,
{\sl Mod. Phys. Lett. } {\bf A5} (1990) 1574; Syracuse University preprint
SU-4228-433
[TPI-MINN-90/31-T] (1990); A.P. Balachandran, W.D. McGlinn, L.
O'Raifeartaigh, S. Sen and R.D. Sorkin,
University of Notre Dame preprint [SU-4228-496] (1991)
\item C. Aneziris, A.P.Balachandran, L.Kauffman and A.M.Srivastava, {\sl
Int.  J. Mod. Phys.}
{\bf A6} (1991) 2519.
\end{enumerate}

\centerline {\bf Figure Captions}

 Figure 1 :A. A completely overlapping string-antistring configuration.
They are on
 top of each other.\hspace{.1in} B. Partial overlap of a string and an
antistring in a plane
              where the overlapped portions have annihilated and are not
shown.
              \hspace{.1in} C. Same as B,
               but the overlapped portions are exhibited even though they
have annihilated.

 Figure 2 : The path representing exchange of two strings.

 Figure 3 : Distortion of path of Fig. 2.

 Figure 4 : String - antistring configurations at various $\tau$ slices of
Fig. 2.

 Figure 5 : Distortion of path of Fig. 3.

 Figure 6 : Distortion of path of Fig. 5.

 Figure 7 : String - antistring configurations at various $\tau$ slices of
Fig. 6.

 Figure 8 : Creation of string at angle $\theta$.

 Figure 9 : Sequence showing annihilation of antistring by two strings.

 Figure 10: Closed path of two strings with one undergoing  a 2$\pi$
rotation.

 Figure 11: A. A string - antistring configuration in which a small portion
of the string
 is created coming off the top of the antistring.\hspace{.1in} B.  Side
view of A.\hspace{.1in}  C. ~Rotation of
 the small portion of the string to restore the final string configuration.

 Figure 12: Same as Figure 11, except earlier in the creation process.

 Figure 13: A. A string - antistring configuration in which a small protion
of the string
 is annihilated from the bottom of the antistring.\hspace{.1in}  B. Side
view of A.\hspace{.1in} C.~Rotation of
 a small portion of the incoming string required to obtain the
configuration in A.

 Figure 14:  Initial configuration for framed strings. Strings and all
 unit vectors lie the in x-y plane.

 Figure 15:   The configuration of a framed string  if a small
  portion of the string is created coming off the top of the antistring.

 Figure 16:  The configuration near the contact point of the string and
antistring with a small portion
 of the string peeled off the top of the antistring in such a way that the
rest of the string can pop off  the top of the antistring
 and result in a smooth configuration of frames.

 Figure 17.  The resulting string configuration after  string 2$'$ is
created with a
 small portion coming off the top of the antistring.

\end{document}